\documentclass[twocolumn,showpacs,superscriptaddress]{revtex4-1}
\usepackage{mymacros}
\usepackage{graphicx}
\usepackage{units}
\usepackage{multirow}
\usepackage{bigstrut}
\usepackage{overpic}
\usepackage{array}
\usepackage{amsfonts}
\usepackage{amsmath}
\usepackage{amssymb}
\usepackage{amsfonts}
\usepackage{verbatim}
\usepackage{footmisc}
\usepackage{color}
\usepackage{flushend}
\usepackage{cuted}
\RequirePackage{lineno}
\begin{document}


\title{\large \bf Measurements of Absolute Hadronic Branching Fractions of the $\lambdacp$ Baryon  }


\author{
  \begin{small}
    \begin{center}
      M.~Ablikim$^{1}$, M.~N.~Achasov$^{9,e}$, X.~C.~Ai$^{1}$,
      O.~Albayrak$^{5}$, M.~Albrecht$^{4}$, D.~J.~Ambrose$^{44}$,
      A.~Amoroso$^{49A,49C}$, F.~F.~An$^{1}$, Q.~An$^{46,a}$,
      J.~Z.~Bai$^{1}$
% ...
      \\
      \vspace{0.2cm}
      (BESIII Collaboration)\\
      \vspace{0.2cm} {\it
        $^{1}$ Institute of High Energy Physics, Beijing 100049, People's Republic of China\\
% ...
$^{4}$ Bochum Ruhr-University, D-44780 Bochum, Germany\\
        $^{5}$ Carnegie Mellon University, Pittsburgh, Pennsylvania 15213, USA\\
% ...
$^{9}$ G.I. Budker Institute of Nuclear Physics SB RAS (BINP), Novosibirsk 630090, Russia\\
% ...
$^{44}$ University of Rochester, Rochester, New York 14627, USA\\
% ...
$^{46}$ University of Science and Technology of China, Hefei 230026, People's Republic of China\\
% ...
$^{49}$ (A)University of Turin, I-10125, Turin, Italy; (B)University of Eastern Piedmont, I-15121, Alessandria, Italy; (C)INFN, I-10125, Turin, Italy\\
% ...
$^{a}$ Also at State Key Laboratory of Particle Detection and Electronics, Beijing 100049, Hefei 230026, People's Republic of China\\
% ...
$^{e}$ Also at the Novosibirsk State University, Novosibirsk, 630090, Russia\\
% ...
      }\end{center}
    \vspace{0.4cm}
\end{small}
}
\affiliation{}

\begin{abstract}
We report the first  measurement of absolute hadronic branching fractions of  $\lambdacp$ baryon at the $\lambdacp\lambdacm$ production threshold, in the 30 years since the $\lambdacp$ discovery.  In total,  twelve Cabibbo-favored $\lambdacp$ hadronic decay modes are analyzed with a double-tag technique, based on a sample of 567\,pb$^{-1}$ of $\ee$ collisions  at $\sqrt{s}=4.599\gev$  recorded with the BESIII detector.
A global least-squares fitter is utilized to improve the measured precision.
Among the measurements for twelve $\lambdacp$ decay modes, the  branching fraction for $\lambdacp \rightarrow pK^-\pi^+$ is determined to be $(5.84\pm0.27\pm0.23)\%$, where the first uncertainty is statistical and the second is systematic.
In addition, the measurements of the branching fractions of the other eleven Cabibbo-favored hadronic decay modes are significantly improved.
\end{abstract}
\maketitle


Charmed baryon decays provide crucial information for the study of both strong and weak interactions. Hadronic  decays of $\lambdacp$, the lightest charmed baryon with quark configuration $udc$,  provide important input to $\Lambda_b$ physics as $\Lambda_b$ decays dominantly to $\lambdacp$~\cite{Dytman:2002yd,Rosner:2012gj}.
Improved measurements of  the $\lambdacp$ hadronic decays can be used to constrain fragmentation functions of charm and bottom quarks by counting inclusive heavy flavor baryons~\cite{Abreu:1999vw}.
Most $\lambdacp$ branching fractions (BF) have until now been obtained by combining measurements of ratios with a single branching fraction of the \emph{golden} reference mode $\lambdacp \rightarrow p K^-\pi^+$, thus introducing strong correlations and compounding uncertainties.
The experimentally averaged BF, $\br{\Lmodeb}=(5.0\pm1.3)\%$~\cite{Agashe:2014kda}, has large uncertainty due to the introduction of model assumptions on $\lambdacp$ inclusive decays in these measurements~\cite{Jaffe:2000nw}. Recently, the Belle experiment reported $\br{\Lmodeb}=(6.84\pm0.24^{+0.21}_{-0.27})\%$ with a precision improved by a factor of 5 over previous results~\cite{Zupanc:2013iki}.
However, most hadronic BFs still have poor precision~\cite{Agashe:2014kda}.
In this Letter, we present the first simultaneous  determination of multiple $\lambdacp$ absolute BFs.

Our analysis is based on a data sample with an integrated luminosity of 567\,pb$^{-1}$~\cite{Ablikim:2015nan} collected with the BESIII detector~\cite{Ablikim:2009aa} at the center-of-mass energy of $\sqrt{s}=4.599$\,GeV.
At this energy,  no additional hadrons accompanying the $\lambdacp\lambdacm$ pairs are produced.
Previously, the Mark III collaboration measured $D$ hadronic BFs at the $D\bar{D}$ threshold using a double-tag technique, which relies on fully reconstructing both $D$ and $\bar{D}$ decays~\cite{mark3}.
This technique obviates the need for knowledge of the luminosity or the production cross section.
We employ a similar technique~\cite{Ablikim:2015prg} using BESIII data near the $\lambdacp\lambdacm$ threshold,  resulting in improved measurements of charge-averaged  BFs for twelve Cabibbo-favored hadronic decay modes: $\lambdacp \to \modea$, $\modeb$, $\modec$, $\moded$, $\modee$, $\modeaa$, $\modebb$, $\modedd$, $\modeaaa$, $\modeccc$, $\modeddd$, and $\modeeee$~\cite{c-omega}. Throughout the Letter, charge-conjugate modes are implicitly assumed, unless otherwise stated.

To identify the $\lambdacp\lambdacm$ signal candidates, we first reconstruct one $\lambdacm$ baryon [called a single tag (ST)] through the final states of any of the twelve modes.
For a given decay mode $j$, the ST yield is determined to be
\begin{eqnarray}
N_{j}^{\rm ST}&=&N_{\lambdacp\lambdacm}\cdot\mathcal{B}_{j}\cdot\varepsilon_{j},
\label{eq:st}
\end{eqnarray}
where $N_{\lambdacp\lambdacm}$ is the total number of produced $\lambdacp\lambdacm$ pairs and $\varepsilon_{j}$ is the corresponding efficiency.
Then we define double-tag (DT) events as those where the partner $\lambdacp$ recoiling against the $\lambdacm$ is reconstructed in one of the twelve modes. That is, in DT events, the $\lambdacp$$\lambdacm$ event is fully reconstructed.   The DT yield with $\lambdacp \to i$ (signal mode) and $\lambdacm \to j$ (tagging mode) is
\begin{eqnarray}
N_{ij}^{\rm DT}&=&N_{\lambdacp\lambdacm}\cdot\mathcal{B}_{i}\cdot\mathcal{B}_{j}\cdot\varepsilon_{ij},
\label{eq:dt}
\end{eqnarray}
where $\varepsilon_{ij}$ is the efficiency for simultaneously reconstructing modes $i$ and $j$.
Hence, the ratio of the DT yield ($N_{ij}^{\rm DT}$) and ST yield ($N_{j}^{\rm ST}$) provides an absolute measurement of the BF:
\begin{eqnarray}
\mathcal{B}_{i} = \frac{N_{ij}^{\rm DT}}{N_{j}^{\rm ST}}  \frac{\varepsilon_{j}}{\varepsilon_{ij}}.
\label{eq:double}
\end{eqnarray}
Because of the large acceptance of the BESIII detector and the low multiplicities of $\Lambda_c$ hadronic  decays, $\varepsilon_{ij} \approx  \varepsilon_{i}\varepsilon_{j}$.
Hence, the ratio $\varepsilon_{j}/\varepsilon_{ij}$ is insensitive to most systematic effects associated with the decay mode $j$, and a signal BF $\mathcal{B}_{i}$ obtained using this procedure is nearly independent of the efficiency of the tagging mode.  Therefore, $\mathcal{B}_{i}$ is sensitive to the signal mode efficiency ($\varepsilon_{i}$), whose uncertainties dominate the contribution to the systematic error from the efficiencies.
According to Eqs.~(\ref{eq:st}) and~(\ref{eq:dt}), the total DT yield with $\lambdacp \to i$ (signal mode) over the twelve ST modes is determined to be
\begin{eqnarray}
N_{i-}^{\rm DT}&=&N_{\lambdacp\lambdacm}\cdot \sum_{j} \mathcal{B}_{i}\cdot\mathcal{B}_{j}\cdot\varepsilon_{i-}^{\rm DT},
\label{eq:tot2}
\end{eqnarray}
where $\varepsilon_{i-}^{\rm DT}\equiv\frac{\sum_{j}(\mathcal{B}_{j}\cdot\varepsilon_{ij})}{\sum_{j}\mathcal{B}_{j}}$ is the average DT efficiency weighted over the twelve modes.

The BESIII detector is an approximately cylindrically symmetric detector with $93\%$ coverage of the solid angle around the $\ee$ interaction point (IP).
The components of the apparatus, ordered by distance from the IP, are a 43-layer small-cell main drift chamber (MDC), a time-of-flight (TOF) system based on plastic scintillators with two layers in the barrel region and one layer in the end-cap region, a 6240-cell CsI(Tl) crystal electromagnetic calorimeter (EMC), a superconducting solenoid magnet providing a 1.0\,T magnetic field aligned with the beam axis, and resistive-plate muon-counter layers interleaved with steel.
The momentum resolution for charged tracks in the MDC is $0.5\%$ for a transverse momentum of $1\gevc$.
The energy resolution in the EMC is $2.5\%$ in the barrel region and $5.0\%$ in the end-cap region for $1\gev$ photons.
Particle identification (PID) for charged tracks combines measurements of the energy deposit d$E$/d$x$ in MDC and flight time in TOF and forms likelihoods $\mathcal{L}(h)~(h=p,K,\pi)$ for a hadron $h$ hypothesis.
More details about the BESIII detector are provided elsewhere~\cite{Ablikim:2009aa}.


High-statistics Monte Carlo (MC) simulations of $\ee$ annihilations are used to understand backgrounds and to estimate detection efficiencies.
The simulation includes the beam-energy spread and initial-state radiation (ISR) of the $\ee$ collisions as simulated with KKMC~\cite{Jadach:2000ir}.
The inclusive MC sample consists of $\lambdacp\lambdacm$ events, $D_{(s)}$ production~\cite{Brambilla:2010cs}, ISR return to lower-mass $\psi$ states, and continuum processes $\ee\to q\bar{q}~(q=u,d,s)$.
Decay modes as specified in the Particle Data Group summary (PDG)~\cite{Agashe:2014kda} are modeled with EVTGEN~\cite{Lange:2001uf}.
For the MC production of $\LLpair$, the observed cross sections are taken into account, and phase-space-generated $\lambdacp$ decays are reweighted according to the observed behaviors in data.
All final tracks and photons are fed into a GEANT4-based~\cite{ref:geant4} detector simulation package.


Charged tracks detected in the MDC must satisfy $|\cos\theta| < 0.93$ (where $\theta$ is the polar angle with respect to the beam direction) and have a distance of closest approach to the IP of less than 10\,cm along the beam axis and less than 1\,cm in the perpendicular plane, except for those used for reconstructing $\Ks$ and $\Lambda$ decays.
Tracks are identified as protons when the PID determines this hypothesis to have the greatest likelihood ($\mathcal{L}(p)>\mathcal{L}(K)$ and $\mathcal{L}(p)>\mathcal{L}(\pi)$), while charged kaons and pions are discriminated based on comparing the likelihoods for these two hypotheses ($\mathcal{L}(K)>\mathcal{L}(\pi)$ or $\mathcal{L}(\pi)>\mathcal{L}(K)$).

Showers in the EMC not associated with any charged track are identified as photon candidates after fulfilling the following requirements.
The deposited energy is required to be larger than 25\,MeV in the barrel ($|\cos\theta|<0.8$) region and 50\,MeV in the end-cap region($0.84<|\cos\theta|<0.92$).
To suppress electronic noise and showers unrelated to the event, the EMC time deviation from the event start time is required to be within (0, 700)\,ns.
The $\pi^{0}$ candidates are reconstructed from photon pairs,
and their invariant masses are required to satisfy $115<M(\gamma\gamma)<150\mevcc$.
To improve momentum resolution, a mass-constrained fit to the $\pi^{0}$ nominal mass is applied to the photon pairs and the resulting energy and momentum of the $\pi^0$ are used for further analysis.

Candidates for $\Ks$ and $\Lambda$ are formed by combining two oppositely charged tracks into the final states $\pip\pim$ and $p\pim$.
For these two tracks, their distances of closest approaches to the IP must be within $\pm$20\,cm along the beam direction.
No distance constraints in the transverse plane are required.
The charged $\pi$ is not subjected to the PID requirements
described above, while proton PID is implemented in order to improve signal significance.
The two daughter tracks are constrained to originate from a common decay vertex by requiring the $\chi^2$ of the vertex fit to be less than 100.
Furthermore, the decay vertex is required to be separated from the IP by a distance of at least twice the fitted vertex resolution.
The fitted momenta of the $\pip\pim$ and $p\pim$ are used in the further analysis.
We impose requirements $487<M(\pi^+\pi^{-})<511\,\mevcc$ and $1111<M(p\pi^{-})<1121\,\mevcc$ to select $\Ks$ and $\Lambda$ signal candidates, respectively, which are within about 3 standard deviations from their nominal masses.
To form $\Sigmazero$, $\Sigmap$ and $\omega$ candidates, requirements on the invariant masses of $1179<M(\Lambda\gm)<1203\mevcc$, $1176<M(p \pizero)<1200\mevcc$ and $760<M(\pi^{+}\pi^{-}\pi^{0})<800\mevcc$, are imposed.

When we reconstruct the decay modes $\modec$, $\moded$ and $\modeddd$, possible backgrounds from $\Lambda\to p \pim$ in the final states are rejected by requiring $M(p\pim)$ outside the range $(1110, 1120)\mevcc$.
In addition, for the mode $\modec$, candidate events within the range $1170<M(p\pi^0)<1200\mevcc$ are excluded to suppress $\Sigma^+$ backgrounds.
To remove $\Ks$ candidates in the modes $\modedd$, $\modeccc$ and $\modeddd$, masses of any pairs of $\pip\pim$ and $\pi^0\pi^0$ are not allowed to fall in the range (480, 520)$\mevcc$.

\begin{figure}[tp!]
\centering
\includegraphics[width=\linewidth]{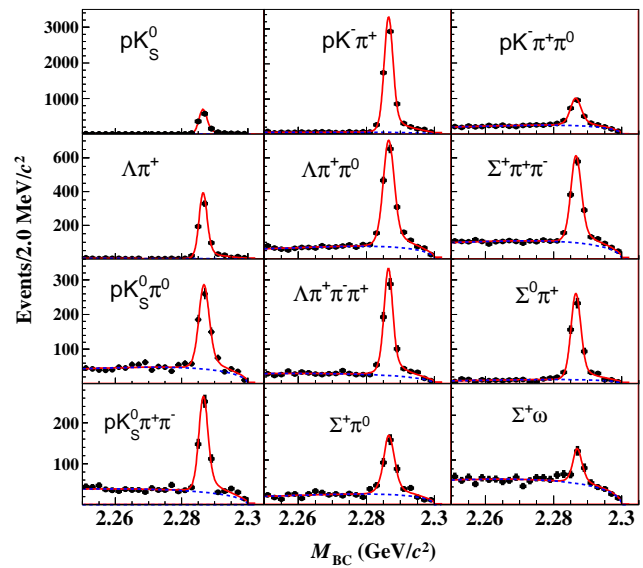}
\caption{Fits to the ST $M_{\rm BC}$ distributions in data for the different decay modes. Points with error bars are data, solid lines are the sum of the fit functions, and dashed lines are the background shapes.}
\label{fig:ST_datafit}
\end{figure}

To discriminate $\Lambda_c$ candidates from background, two variables reflecting energy and momentum conservation are used.
First, we calculate the energy difference, $\Delta{}E \equiv E - E_{\rm beam}$, where $E$ is the total measured energy of the $\Lambda_c$ candidate and $E_{\rm beam}$ is the average value of the $e^+$ and $e^-$ beam energies.
For each tag mode, candidates are rejected if they fail the $\Delta{}E$ requirements in Table~\ref{tab:STyields}, which correspond to about 3 times the resolutions.
Second, we define the beam-constrained mass $M_{\rm BC}$ of the $\Lambda_c$ candidates by substituting the beam-energy $E_{\rm beam}$ for the energy $E$ of the $\Lambda_c$ candidates,
$M_{\rm BC}c^2\equiv \sqrt{E_{\rm beam}^2-p^2c^2}$,
where $p$ is the measured $\Lambda_c$ momentum in the center-of-mass system of the $\ee$ collision.
Figure~\ref{fig:ST_datafit} shows the $M_{\rm BC}$ distributions for the ST samples, where evident $\Lambda_c$ signals peak at the nominal $\Lambda_c$ mass position (2286.46$\pm$0.14) MeV/$c^2$~\cite{Agashe:2014kda}.
The MC simulations show that peaking backgrounds and cross feeds among the twelve ST modes are negligible.

We perform unbinned extended  maximum likelihood fits to the $M_{\rm BC}$ distributions to obtain the ST yields, as illustrated in Fig.~\ref{fig:ST_datafit}.
In each fit, the signal shape is derived from MC simulations of the signal ST modes convolved with a Gaussian function to account for imperfect modeling of the detector resolution and beam-energy spread. The parameters of the Gaussians are allowed to vary in the fits.
Backgrounds for each mode are described with the ARGUS function~\cite{ref:argus}.
The resultant ST yields in the signal region $2276<M_{\rm BC}<2300\mevcc$ and the corresponding detection efficiencies are listed in Table~\ref{tab:STyields}.

\begin{table}[tp!]
  \begin{center}
 \caption{Requirement on $\Delta{}E$, ST yields, DT yields and detection efficiencies for each of the decay modes. The uncertainties are statistical only. The quoted efficiencies do not include any subleading BFs.}
  \resizebox{\linewidth}{!}{
  \begin{tabular}{l|c|c|c|c|c}
      \hline \hline
 \rule{0pt}{9pt}  Mode  &$\Delta{}E$ (MeV) & $N_{j}^{\rm ST}$ & $\varepsilon_{j}(\%)$ & $N_{i-}^{\rm DT}$ & $\varepsilon_{i-}^{\rm DT}(\%)$ \\ \hline
  $\textbf{$\modea$}$   & $(-20,20)$  & $1243\pm37$ & $55.9$ & $97\pm10$ & $16.6$ \\
$\textbf{$\modeb$}$   & $(-20,20)$  & $6308\pm88$ & $51.2$ & $420\pm22$ & $14.1$ \\
$\textbf{$\modec$}$   & $(-30,20)$  & $558\pm33$ & $20.6$ & $47\pm8$ & $6.8$ \\
$\textbf{$\moded$}$   & $(-20,20)$  & $485\pm29$ & $21.4$ & $34\pm6$ & $6.4$ \\
$\textbf{$\modee$}$   & $(-30,20)$  & $1849\pm71$ & $19.6$ & $176\pm14$ & $7.6$ \\
$\textbf{$\modeaa$}$  & $(-20,20)$  & $706\pm27$ & $42.2$ & $60\pm8$ & $12.7$ \\
$\textbf{$\modebb$}$  & $(-30,20)$  & $1497\pm52$ & $15.7$ & $101\pm13$ & $5.4$ \\
$\textbf{$\modedd$}$  & $(-20,20)$  & $609\pm31$ & $12.0$ & $53\pm7$ & $3.6$ \\
$\textbf{$\modeaaa$}$ & $(-20,20)$  & $522\pm27$ & $29.9$ & $38\pm6$ & $9.9$ \\
$\textbf{$\modeccc$}$ & $(-50,30)$  & $309\pm24$ & $23.8$ & $25\pm5$ & $8.0$ \\
$\textbf{$\modeddd$}$ & $(-30,20)$  & $1156\pm49$ & $24.2$ & $80\pm9$ & $8.1$ \\
$\textbf{$\modeeee$}$ & $(-30,20)$  & $157\pm22$ & $9.9$ & $13\pm3$ & $3.8$ \\
\hline  \hline
   \end{tabular}
      }
      \label{tab:STyields}
  \end{center}
  \end{table}

\begin{figure}[tp!]
\centering
\includegraphics[width=\linewidth]{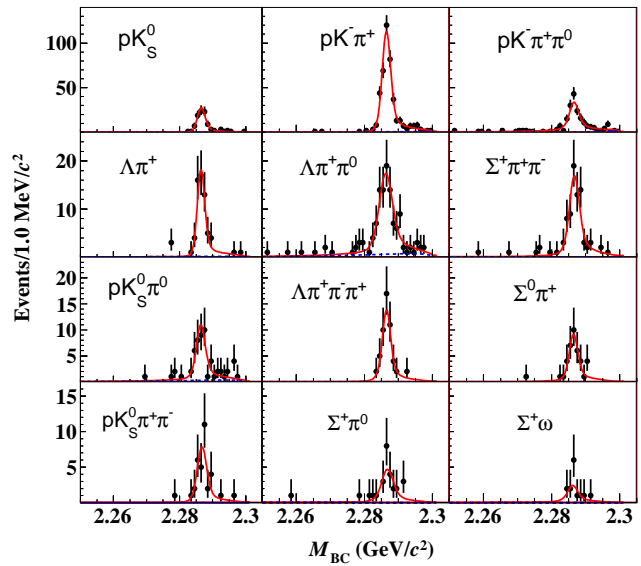}
\caption{Fits to the DT $M_{\rm BC}$ distributions in data for different signal modes. Points with error bars are data, solid lines are the sum of fit functions, and dashed lines are background shapes.}
\label{fig:DT_datafit}
\end{figure}

In the signal candidates of the twelve ST modes, a specific mode $\lambdacp\to i$ is formed from the remaining tracks and showers recoiling against the ST $\lambdacm$.
We combine the DT signal candidates over the twelve ST modes and plot the distributions of the  $M_{\rm BC}$ variable in  Fig.~\ref{fig:DT_datafit}.
We follow the same fit strategy as in the ST samples to estimate the total DT yield $N_{i-}^{\rm DT}$ in Eq.~\eqref{eq:tot2}, except that the DT signal shapes are derived from the DT signal MC samples and convolved with the Gaussian function.  The parameters of the Gaussians are also allowed to vary in the fits. The extracted DT yields are listed in Table~\ref{tab:STyields}. The $12\times12$ DT efficiencies $\varepsilon_{ij}$ are evaluated based on the DT signal MC samples, in order to extract the BFs.

Main sources of systematic uncertainties related to the measurement of BFs include tracking, PID, reconstruction of intermediate states and intermediate BFs.
For the $\Delta E$ and $M_{\rm BC}$ requirements, the uncertainties are negligible, as we correct resolutions in MC samples to accord with those in data.
Uncertainties associated with the efficiencies of the tracking and PID of charged particles are estimated by studying a set of control samples of $\ee\to \pip\pip\pim\pim$, $K^+K^-\pi^+\pi^-$ and $p\bar{p}\pi^+\pi^-$ based on data taken at energies above $\sqrt{s}=4.0$\,GeV.
An uncertainty of 1.0\% is  assigned to each $\pi^0$ due to the reconstruction efficiency.
The uncertainties of detecting $\Ks$ and $\Lambda$ are determined to be 1.2\% and 2.5\%, respectively.
Reweighting factors for the twelve signal models are varied within their statistical uncertainties obtained from the ST data samples.
Deviations of the resultant efficiencies are taken into account in systematic uncertainties.
Systematic uncertainties due to limited statistics in MC samples are included.
Uncertainties on the BFs of intermediate state decays from the PDG~\cite{Agashe:2014kda} are also included.
A summary of systematic uncertainties are given in Table~\ref{tab:sys_err}.

\begin{table}[tp]
  \begin{center}
	\caption{Summary of systematic uncertainties, in percent. The total numbers are derived from the least-squares fit, by taking into account correlations among different modes.}
	\resizebox{\linewidth}{!}{
    \begin{tabular}{l|cccccccc|c}
      \hline \hline
      \multirow{2}{*}{Source} & \multirow{2}{*}{Tracking}  &  \multirow{2}{*}{PID}  & \multirow{2}{*}{$\Ks$}  & \multirow{2}{*}{$\Lambda$} & \multirow{2}{*}{$\pizero$}  & Signal &MC          & Quoted   &  \multirow{2}{*}{Total} \\
                             &                             &                        &                         &                            &                             &   model                     & stat.   &BFs  &      \\ \hline
      $\textbf{$\modea$}$   & 1.3   & 0.3   & 1.2    &      &       & 0.2 & 0.4 &  0.1&  2.0 \\
      $\textbf{$\modeb$}$   & 2.5   & 3.2   &        &      &       &     & 0.2 &     &  3.9 \\
      $\textbf{$\modec$}$   & 1.1   & 1.6   & 1.2    &      &  1.0  & 1.0 & 0.5 &  0.1&  2.7 \\
      $\textbf{$\moded$}$   & 2.8   & 5.4   & 1.2    &      &       & 0.5 & 0.5 &  0.1&  5.9 \\
      $\textbf{$\modee$}$   & 3.3   & 5.8   &        &      &  1.0  & 2.0 & 0.5 &     &  6.6 \\
      $\textbf{$\modeaa$}$  & 1.0   & 1.0   &        & 2.5  &       & 0.5 & 0.5 &  0.8&  2.4 \\
      $\textbf{$\modebb$}$  & 1.0   & 1.0   &        & 2.5  &  1.0  & 0.6 & 0.6 &  0.8&  2.7 \\
      $\textbf{$\modedd$}$  & 3.0   & 3.0   &        & 2.5  &       & 0.8 & 0.8 &  0.8&  4.7 \\
      $\textbf{$\modeaaa$}$ & 1.0   & 1.0   &        & 2.5  &       & 1.7 & 0.7 &  0.8&  2.4 \\
      $\textbf{$\modeccc$}$ & 1.3   & 0.3   &        &      &  2.0  & 1.7 & 0.8 &  0.1&  2.5 \\
      $\textbf{$\modeddd$}$ & 3.0   & 3.7   &        &      &  1.0  & 0.8 & 0.4 &  0.1&  4.7 \\
      $\textbf{$\modeeee$}$ & 3.0   & 3.2   &        &      &  2.0  & 7.1 & 1.0 &  0.8&  4.5 \\
      \hline\hline
    \end{tabular}
}\label{tab:sys_err}
  \end{center}
  \end{table}

We use a least-squares fitter, which considers statistical and systematic correlations among the different hadronic modes, to obtain the BFs of the twelve $\lambdacp$ decay modes globally.
Details of this fitter are discussed in Ref.~\cite{Guan:2013hua}.
In the fitter, the precisions of the twelve  BFs are constrained to a common variable, $N_{\lambdacp\lambdacm}$, according to Eqs.~\eqref{eq:st} and ~\eqref{eq:tot2}.
In total, there are thirteen free parameters (twelve $\mathcal{B}_{i}$ and $N_{\lambdacp\lambdacm}$) to be estimated.
As peaking backgrounds in ST modes and cross feeds among the twelve ST modes are suppressed to a negligible level, they are not considered in the fit.

The extracted BFs of $\lambdacp$ are listed  in Table~\ref{tab:compare};
the correlation matrix is available in the Supplemental Material.
The total number of $\lambdacp\lambdacm$ pairs produced is obtained to be $N_{\lambdacp\lambdacm}=(105.9\pm4.8\pm0.5)\times10^3$.
The goodness-of-fit is evaluated as $\chi^2/\text{ndf}=9.9/(24-13)=0.9$.

\begin{table}[tp!]
  \begin{center}
   \caption{Comparison of the measured BFs in this work with previous results from PDG~\cite{Agashe:2014kda}. For our results, the first uncertainties are statistical and the second are systematic.
 }
    \begin{tabular}{l|c|c}
      \hline \hline
  Mode  & This work (\%) & PDG (\%)  \\ \hline
$\textbf{$\modea$}$   & $1.52\pm0.08\pm0.03$ & $1.15\pm0.30$   \\
$\textbf{$\modeb$}$   & $5.84\pm0.27\pm0.23$ & $5.0\pm1.3$  \\
$\textbf{$\modec$}$   & $1.87\pm0.13\pm0.05$ & $1.65\pm0.50$   \\
$\textbf{$\moded$}$   & $1.53\pm0.11\pm0.09$ & $1.30\pm0.35$    \\
$\textbf{$\modee$}$   & $4.53\pm0.23\pm0.30$ & $3.4\pm1.0$   \\
$\textbf{$\modeaa$}$  & $1.24\pm0.07\pm0.03$ & $1.07\pm0.28$   \\
$\textbf{$\modebb$}$  & $7.01\pm0.37\pm0.19$ & $3.6\pm1.3$   \\
$\textbf{$\modedd$}$  & $3.81\pm0.24\pm0.18$ & $2.6\pm0.7$   \\
$\textbf{$\modeaaa$}$ & $1.27\pm0.08\pm0.03$ & $1.05\pm0.28$   \\
$\textbf{$\modeccc$}$ & $1.18\pm0.10\pm0.03$ & $1.00\pm0.34$   \\
$\textbf{$\modeddd$}$ & $4.25\pm0.24\pm0.20$ & $3.6\pm1.0$   \\
$\textbf{$\modeeee$}$ & $1.56\pm0.20\pm0.07$ & $2.7\pm1.0$   \\
\hline \hline
   \end{tabular}
     \label{tab:compare}
 \end{center}
\end{table}

To summarize, twelve Cabibbo-favored $\lambdacp$ decay rates are measured by employing a double-tag technique, based on a sample of threshold data at $\sqrt{s}=4.599\gev$ collected at BESIII.
This is the first absolute measurement of the $\lambdacp$ decay branching fractions at the $\lambdacp\lambdacm$ production threshold, in the 30 years since the $\lambdacp$ discovery.
A comparison with previous results is presented in Table~\ref{tab:compare}.
For the golden mode $\br{\modeb}$, our result is consistent with that in PDG, but lower than Belle's with a significance of about $2\sigma$.
For the branching fractions of the other modes, the precisions are improved by factors of $3\sim6$ compared to the world average values.

The BESIII Collaboration thanks the staff of BEPCII and the IHEP computing center for their strong support.
This work is supported in part by National Key Basic Research Program of China under Contract No. 2015CB856700;
National Natural Science Foundation of China (NSFC) under Contracts No. 11125525, No. 11235011, No. 11275266, No. 11322544, No. 11335008 and No. 11425524;
the Chinese Academy of Sciences (CAS) Large-Scale Scientific Facility Program;
the CAS Center for Excellence in Particle Physics (CCEPP);
the Collaborative Innovation Center for Particles and Interactions (CICPI); Joint Large-Scale Scientific Facility Funds of the NSFC and CAS under Contracts No. 11179007, No. U1232201 and No. U1332201;
CAS under Contracts No. KJCX2-YW-N29 and No.  KJCX2-YW-N45; 100 Talents Program of CAS;
National 1000 Talents Program of China;
INPAC and Shanghai Key Laboratory for Particle Physics and Cosmology;
German Research Foundation DFG under Contract No. Collaborative Research Center CRC-1044;
Istituto Nazionale di Fisica Nucleare, Italy;
Koninklijke Nederlandse Akademie van Wetenschappen (KNAW) under Contract No. 530-4CDP03;
Ministry of Development of Turkey under Contract No. DPT2006K-120470;
National Natural Science Foundation of China (NSFC) under Contracts No. 11405046 and No.  U1332103;
Russian Foundation for Basic Research under Contract No. 14-07-91152;
The Swedish Resarch Council;
U. S. Department of Energy under Contracts No. DE-FG02-04ER41291, No. DE-FG02-05ER41374, No. DE-SC0012069, and No. DESC0010118;
U.S. National Science Foundation;
University of Groningen (RuG) and the Helmholtzzentrum fuer Schwerionenforschung GmbH (GSI), Darmstadt;
and WCU Program of National Research Foundation of Korea under Contract No. R32-2008-000-10155-0.


\newpage

\section*{Supplemental Material}  
We present the correlation matrix of the branching fraction fit. In total, there are thirteen correlated items; one $N_{\lambdacp\lambdacm}$ and twelve branching fractions.
\begin{table*}[!h]
  \begin{center}
  \footnotesize
 \caption{Correlation coefficients among thirteen fit parameters, including both statistical and systematic uncertainties.}
  \resizebox{\linewidth}{!}{
  \begin{tabular}{lccccccccccccc}
      \hline \hline
                     & $N_{\lambdacp\lambdacm}$ & $\mathcal{B}(\textbf{$\modea$})$ & $\mathcal{B}(\textbf{$\modeb$})$   & $\mathcal{B}(\textbf{$\modec$})$ & $\mathcal{B}(\textbf{$\moded$})$ & $\mathcal{B}(\textbf{$\modee$})$ & $\mathcal{B}(\textbf{$\modeaa$})$ & $\mathcal{B}(\textbf{$\modebb$})$ & $\mathcal{B}(\textbf{$\modedd$})$ & $\mathcal{B}(\textbf{$\modeaaa$})$ & $\mathcal{B}(\textbf{$\modeccc$})$ & $\mathcal{B}(\textbf{$\modeddd$})$ & $\mathcal{B}(\textbf{$\modeeee$})$   \\ \hline
$N_{\lambdacp\lambdacm}$              & $ 1   $ & $-0.80$ & $-0.71$ & $-0.55$ & $-0.42$ & $-0.43$ & $-0.68$ & $-0.64$ & $-0.48$ & $-0.57$ & $-0.46$ & $-0.51$ & $-0.21$ \\
$\mathcal{B}(\textbf{$\modea$})$      &         & $ 1   $ & $ 0.69$ & $ 0.52$ & $ 0.47$ & $ 0.47$ & $ 0.59$ & $ 0.56$ & $ 0.50$ & $ 0.51$ & $ 0.41$ & $ 0.54$ & $ 0.23$ \\
$\mathcal{B}(\textbf{$\modeb$})$      &         &         & $ 1   $ & $ 0.57$ & $ 0.73$ & $ 0.84$ & $ 0.64$ & $ 0.61$ & $ 0.70$ & $ 0.54$ & $ 0.42$ & $ 0.80$ & $ 0.37$ \\
$\mathcal{B}(\textbf{$\modec$})$      &         &         &         & $ 1   $ & $ 0.42$ & $ 0.47$ & $ 0.43$ & $ 0.43$ & $ 0.40$ & $ 0.37$ & $ 0.32$ & $ 0.47$ & $ 0.21$ \\
$\mathcal{B}(\textbf{$\moded$})$      &         &         &         &         & $ 1   $ & $ 0.70$ & $ 0.42$ & $ 0.42$ & $ 0.54$ & $ 0.36$ & $ 0.26$ & $ 0.63$ & $ 0.29$ \\
$\mathcal{B}(\textbf{$\modee$})$      &         &         &         &         &         & $ 1   $ & $ 0.46$ & $ 0.47$ & $ 0.61$ & $ 0.40$ & $ 0.30$ & $ 0.74$ & $ 0.35$ \\
$\mathcal{B}(\textbf{$\modeaa$})$     &         &         &         &         &         &         & $ 1   $ & $ 0.65$ & $ 0.57$ & $ 0.57$ & $ 0.34$ & $ 0.49$ & $ 0.21$ \\
$\mathcal{B}(\textbf{$\modebb$})$     &         &         &         &         &         &         &         & $ 1   $ & $ 0.56$ & $ 0.57$ & $ 0.36$ & $ 0.50$ & $ 0.22$ \\
$\mathcal{B}(\textbf{$\modedd$})$     &         &         &         &         &         &         &         &         & $ 1   $ & $ 0.50$ & $ 0.28$ & $ 0.59$ & $ 0.27$ \\
$\mathcal{B}(\textbf{$\modeaaa$})$    &         &         &         &         &         &         &         &         &         & $ 1   $ & $ 0.28$ & $ 0.42$ & $ 0.18$ \\
$\mathcal{B}(\textbf{$\modeccc$})$    &         &         &         &         &         &         &         &         &         &         & $ 1   $ & $ 0.34$ & $ 0.16$ \\
$\mathcal{B}(\textbf{$\modeddd$})$    &         &         &         &         &         &         &         &         &         &         &         & $ 1   $ & $ 0.33$ \\
$\mathcal{B}(\textbf{$\modeeee$})$    &         &         &         &         &         &         &         &         &         &         &         &         & $ 1   $ \\
 \hline
   \end{tabular}
   }
   \label{tab:correlation}
 \end{center}
\end{table*}


\begin{thebibliography}{99}
\bibitem{c-omega}
The $\Sigma^+\pi^+\pi^-\pi^0$ mode suffers from huge combinatorial backgrounds. We restrict $\pi^+\pi^-\pi^0$ from $\omega$ decays and only measure the resonant component $\Sigma^+\omega$, where the non-resonant final states $\Sigma^+\pi^+\pi^-\pi^0$ are negligible.
\end{thebibliography}
\end{document}